\documentclass[runningheads,orivec]{llncs}

\usepackage[T1]{fontenc}
\usepackage{amsmath}
\usepackage{xcolor}
\usepackage{graphicx} 
\usepackage{hyperref}
\usepackage{url}
\usepackage[dvipsnames]{xcolor}
\definecolor{gainsboro}{rgb}{0.86,0.86,0.86}

\usepackage{multirow}
\usepackage{fancybox}

\usepackage{listings}
\lstdefinelanguage{Kconfig}{
    morestring=[b]",
    morestring=[d]’,
    morecomment=[l]{\#},
    commentstyle=\itshape\color{brown},
    backgroundcolor=\color{white},
    keywordstyle=\color{magenta}\bfseries,
    numberstyle=\tiny\color{gray},
    stringstyle=\color{ForestGreen}\ttfamily,
    basicstyle=\ttfamily\scriptsize,
    breakatwhitespace=false,
    emph={if, choice,  prompt, depends, on, option, env, comment, visible, help,  mainmenu, menu, endif, endchoice, endmenu, endmainmenu, default},
    emphstyle={\color{Blue}\bfseries},
    classoffset=1, 
    otherkeywords={y, n},
    morekeywords={y, n},
    keywordstyle=\color{brown}\bfseries,
    classoffset=2, 
    otherkeywords={bool, tristate, string, int, hex, config},
    morekeywords={bool, tristate, string, int, hex, config},
    keywordstyle=\color{purple}\bfseries,
    classoffset=3, 
    otherkeywords={\&, |, =, ==, !=, !},
    morekeywords={\&, |, =, ==, !=, !},
    keywordstyle=\color{yellow}\bfseries,
    classoffset=4, 
    keywordstyle=\color{cyan}\bfseries,
    classoffset=5, 
    otherkeywords={source},
    morekeywords={source},
    keywordstyle=\color{magenta}\bfseries,
    classoffset=6, 
    otherkeywords={select},
    morekeywords={select},
    keywordstyle=\color{YellowGreen}\bfseries,
    classoffset=7, 
    otherkeywords={range},
    morekeywords={range},
    keywordstyle=\color{brown}\bfseries,
    classoffset=0,
    breaklines=true,
    captionpos=b,
    keepspaces=true,
    numbers=left,
    numbersep=8pt,
    showspaces=false,
    showstringspaces=false,
    showtabs=false,
    tabsize=2
}

\lstdefinelanguage{graphs}{
    morecomment=[l]{\#},
    commentstyle=\itshape\color{brown},
    backgroundcolor=\color{white},
    keywordstyle=\color{magenta}\bfseries,
    numberstyle=\tiny\color{gray},
    stringstyle=\color{ForestGreen}\ttfamily,
    basicstyle=\ttfamily\scriptsize,
    breakatwhitespace=false,
    emph={Vertices,  Arcs, Edges},
    emphstyle={\color{Blue}\bfseries},
    classoffset=1, 
    otherkeywords={y, n},
    morekeywords={y, n},
    keywordstyle=\color{brown}\bfseries,
    classoffset=2, 
    otherkeywords={bool, tristate, string, int, hex, config},
    morekeywords={bool, tristate, string, int, hex, config},
    keywordstyle=\color{purple}\bfseries,
    classoffset=3, 
    otherkeywords={\&, |, =, ==, !=, !},
    morekeywords={\&, |, =, ==, !=, !},
    keywordstyle=\color{yellow}\bfseries,
    classoffset=4, 
    keywordstyle=\color{cyan}\bfseries,
    classoffset=5, 
    otherkeywords={source},
    morekeywords={source},
    keywordstyle=\color{magenta}\bfseries,
    classoffset=6, 
    otherkeywords={select},
    morekeywords={select},
    keywordstyle=\color{YellowGreen}\bfseries,
    classoffset=7, 
    otherkeywords={range},
    morekeywords={range},
    keywordstyle=\color{brown}\bfseries,
    classoffset=0,
    breaklines=true,
    captionpos=b,
    keepspaces=true,
    numbers=left,
    numbersep=8pt,
    showspaces=false,
    showstringspaces=false,
    showtabs=false,
    tabsize=2
}

\usepackage{dirtree}
\definecolor{aurometalsaurus}{rgb}{0.43, 0.5, 0.5}

\usepackage{todonotes}

\newcommand{\review}[1]{#1}

\usepackage[noend,]{algpseudocode}
\usepackage{algorithm,algorithmicx}

\algnewcommand{\LineComment}[1]{\Statex \(\triangleright\) #1}

\begin{document}
\title{Reasoning About Variability Models\\Through Network Analysis}
\titlerunning{Network Analysis for Variability Models}
\author{
Jose Manuel Sanchez\inst{1}\orcidID{0009-0004-0436-5725} \and\\
Miguel Angel Olivero\inst{1}\orcidID{0000-0002-6627-3699} \and\\
Ruben Heradio\inst{2}\orcidID{0000-0002-7131-0482} \and\\
Luis Cambelo\inst{2}\orcidID{0009-0004-8160-1365} \and\\
David Fernandez-Amoros\inst{2}\orcidID{0000-0003-3758-0195}}
\authorrunning{J.M. Sanchez et al.}
\institute{
Universidad de Sevilla, Sevilla (Spain)\\
\email{\{jsanchez7,molivero\}@us.es}
\and
Universidad Nacional de Educación a Distancia (UNED), Madrid (Spain)\\
\email{\{rheradio,lcambelo1,david\}@issi.uned.es}
}





%
\maketitle              
\begin{abstract}
Feature models are widely used to capture the configuration space of software systems.
Although automated reasoning has been studied for detecting problematic features and supporting configuration tasks, significantly less attention has been given to the systematic study of the structural properties of feature models at scale.
The approach fills this gap by examining the models' structure through a network analysis perspective.
We focus on three Research Questions concerning
(i) the structural patterns exhibited by these graphs,
(ii) the extent to which such patterns vary across domains and model sources, and
(iii) the usefulness of network‑based indicators for understanding, maintaining, and evolving variability models.
To answer these questions, we analyze a dataset of 5,709 models from 20 repositories, spanning multiple application domains and varying sizes (ranging from 99 to 35,907 variables on their Boolean translation).
To do so, graphs of transitive dependencies and conflicts between features are computed.
Our results reveal consistent structural traits (e.g., the predominance of dependency relations, the presence of highly central features, or characteristic node degree distributions) as well as notable domain‑specific deviations.
These findings ease the identification of maintenance‑relevant features, opportunities for modular decomposition, and indicators of structural fragility.
This approach provides a scalable, graph‑based foundation for the empirical analysis of variability models and contributes quantitative evidence to support future research on their structure and evolution.
\keywords{variability models  \and configurable systems \and network analysis}
\end{abstract}

\vspace{-0.8cm}
\section{Introduction}\label{sec:introduction}
    Over the last decades, feature modeling techniques have been studied in academia and introduced in industrial contexts.
    \textit{Software Product Lines} (SPLs) are a widely adopted approach to systematically manage variability in families of related software products that share assets and features~\cite{clements2002}.
    \textit{Feature Models} (FMs)~\cite{felfernig2024feature} are a common means to capture this variability: they describe the set of valid products as configurations of features that respect a hierarchy and a set of cross-tree constraints.
    As reported by Berger et al.~\cite{berger2013survey}, SPL engineering practices and feature modeling have been adopted in a broad range of domains, including automotive, energy, enterprise systems, eCommerce, aerospace, defense, and medical devices.
    They also highlight that feature modeling is perceived as valuable for managing existing variability, configuring products, specifying requirements, and deriving products.

    The analysis of FMs is crucial to support activities such as maintenance, evolution, and performance optimization~\cite{batory2006automated,benavides2010automated}.
    However, as FMs grow in size and complexity, understanding their global structure becomes increasingly challenging.
    This raises the need for complementary analysis techniques that provide high-level, structural insights into FMs.

    In this work, we explore the application of \textit{network analysis}~\cite{brandes2005network} to perform automated analysis of FMs.
    The key idea is to view FMs as graphs that encode \textit{transitive strong} relationships between features. Here, ``transitive'' refers to relationships that can be formed by chaining other relationships, while ``strong'' indicates that these relationships hold in all configurations that adhere to the feature model.
    To this end, we rely on a representation in which each feature is a node, and require/exclude relationships between features are captured as require arcs/conflict edges, respectively.
    These graphs form a data structure that enables reasoning about reachability, influence, and structural patterns among features.

    Building on prior work on automated FM analysis~\cite{benavides2010automated,heradio2016augmenting,heradio2011supporting}, 4 artifacts are provided for each FM: a \textit{dependency graph}, a \textit{conflict graph}, and two lists containing the \textit{core} and \textit{dead} features of the model (i.e., features that appear in every valid configuration and features that do not appear in any valid configuration, respectively).
    Graph representations abstract away from the syntactic details of the original FMs and preserves information, which is highly relevant to configuration, maintenance, and evolution tasks.
    They also enable the application of a wide range of network analysis techniques to study the structure of FMs.

    We systematically apply network analysis to a large corpus of FMs.
    We consider 5{,}709 FMs, available as an open dataset\footnote{\url{https://doi.org/10.5281/zenodo.17790234}}, deriving for each of them its dependency graph, conflict graph, and sets of core and dead features.
    On top of these graph-based representations, we compute and analyze structural metrics to characterize how real-world FMs are organized.
    Our goal is to investigate which patterns are frequent across models and what they suggest about the robustness, configurability, and maintainability of highly configurable systems.
    Specifically, this work addresses the following research questions:
    \begin{itemize}
      \item \textbf{RQ1:} What structural patterns emerge in the dependency and conflict graphs of real-world FMs?
      \item \textbf{RQ2:} How do these structural patterns vary across application domains and model sources?
      \item \textbf{RQ3:} How can network-based indicators derived from these structural patterns support practitioners in understanding, maintaining, and evolving large FMs?
    \end{itemize}

    The remainder of this article is structured as follows.
    Section~\ref{sec:related-work} reviews feature modeling, automated analysis of FMs, and prior work on graph-based and network-based analyses.
    Section~\ref{sec:example} presents two FMs of different sizes to illustrate the difficulty of understanding large models and to motivate the need for network-based analysis.
    Section~\ref{sec:how-to} summarizes the process used to obtain the graph representations.
    Section~\ref{sec:experimental-validation} describes the empirical study conducted on 5,709 FMs, detailing the selected network metrics, discussing the results, and summarizing the main findings.
    Finally, Section~\ref{sec:conclusions} outlines the implications of our results and sketches directions for future work.
\vspace{-2mm}
\section{Background}\label{sec:related-work}

This section recalls basic notions on feature models and automated analysis, introduces the semantic relations that underlie strong graphs, and briefly situates our work in the context of network analysis applied to FMs.
\vspace{-2mm}
    \subsection{Feature models and automated analysis}
    \label{subsec:fms-automated-analysis}

    A FM defines a family of products as configurations of features that satisfy both a feature hierarchy (mandatory and optional features, alternative and or-groups) and a set of constraints.
    Typical FM analyses include detecting anomalies (e.g., void models, dead features, false optional features), counting valid configurations, and supporting configuration and testing tasks. A common approach is to encode FMs as propositional formulas and utilize SAT or CSP solvers to answer these analysis queries~\cite{liang2015sat}.

    \vspace{-2mm}
    \subsection{Indirect relations and transitive strong graphs}
    \label{subsec:strong-graphs}

    While the \textit{direct} require and exclude constraints between features are defined, FM semantics produce numerous additional \textit{indirect} relationships that often go unnoticed. The interplay of hierarchy, group relations, and cross-tree constraints can enforce situations in which one feature must always be selected if another is chosen, or it can prohibit two features from co-occurring in any valid configuration. Understanding these semantic relationships is crucial for managing FMs, as they can highlight unintentionally coupled features or reveal hidden conflicts~\cite{benavides2010automated,heradio2016augmenting,heradio2011supporting}.

    In this work, a feature $f$ is said to have a \emph{strong dependency} on a feature $g$ if, in every valid configuration that conforms to the FM, whenever $f$ is selected, then $g$ is also selected.
    Two features $f$ and $g$ are in \emph{strong conflict} if they never appear together in any valid configuration.
    Note that our definition involves \textit{transitivity}: a strong relationship takes into account chaining intermediate relationships specified in the FM (e.g., if according to the FM, $f$ depends on $g$, and $g$ depends on $h$, then there is a strong dependency between $f$ and $h$).

    Strong relations can be organized as graphs.
    In a \textit{dependency graph}, nodes correspond to features and there is a directed arc $f \rightarrow g$ whenever $f$ strongly depends on $g$.
    In a \textit{conflict graph}, nodes are again features and there is an undirected edge between $f$ and $g$ whenever they are in strong conflict.
    Throughout the paper, the term \emph{strong graphs} refers to both dependency and conflict graphs.
    To keep graphs informative and manageable, trivial or redundant relations are removed (e.g., self-loops or arcs from every node to each core feature).
    Section~\ref{sec:how-to} describes how the graphs are computed.

    \vspace{-2mm}
    \subsection{Network analysis for software and FMs}
    \label{subsec:network-analysis}

    Network analysis provides metrics to study graphs that represent complex systems.
    These metrics have been used in software engineering to analyze call graphs, dependency networks, or fault propagation, with the aim of identifying hotspots, critical components, or structurally fragile regions.

    In variability modeling, graph-based representations of FMs have been used to define metrics and assess structural properties, such as commonality and the presence of anomalies.
    Most existing work, however, focuses on logical analysis (e.g., satisfiability, anomaly detection), does not explore structural properties, and does not analyze a large set of FMs~\cite{10.1145/3442389,10.1145/2580950,6980213}.
    \review{Boender~\cite{Boender11} applies graph-based representations to study and analyze the dependencies between packages in real Linux distributions (Debian and Mandriva) and Eclipse.
    The development of directed graphs that represent strong dependencies allows for a structural analysis of software distributions.
    This way, his work proposes a formal model of open-source software distributions with methods to improve quality management.
    Software product lines and feature models are mentioned in the conclusions of Boender's study, regarding an open way for the re-use and extension of his results in the field of software product lines.}

    In contrast to \review{the mentioned existing studies}, this work systematically applies standard network metrics to strong dependency and conflict graphs derived from a large corpus of real-world FMs.
    By treating features as nodes and strong relations as arcs and edges, we compute structural indicators over these graphs and relate them to properties of the underlying models and their domains, complementing traditional logical analyses with a structural perspective.

\vspace{-2mm}
\section{Two motivational examples}
\label{sec:example}

    This section presents two examples that illustrate how the complexity of variability analysis changes with the size of the models.
    We first use a small excerpt of the \textsf{coreboot 4.13}\footnote{\url{https://www.coreboot.org/}} FM to illustrate that even the simplest models contain indirect relationships that are difficult to identify without the use of strong graphs.
    A second example featuring a larger FM, \textsf{Linux 2.6.9 X86 64-bit}\footnote{\url{https://www.kernel.org/}}, shows how strong graphs and their associated metrics become essential when visual inspection is no longer practical.

    \vspace{-2mm}
    \subsection{A tiny FM that hides some subtle indirect relationships}
    Figure~\ref{lst:kconfig-code} shows a fragment of FM of \textsf{coreboot 4.13}, an open-source firmware for computers and embedded systems, which comes from \texttt{src/device/Kconfig} and deals with variability in graphics initialization and framebuffer configuration.
    Each \texttt{config} entry declares a Boolean feature and its dependencies, while \texttt{choice} blocks express mutually exclusive alternatives, ensuring that exactly one option is selected.

\begin{figure}
    \begin{lstlisting}[language=Kconfig]
config HAVE_VGA_TEXT_FRAMEBUFFER
  bool
  depends on !NO_GFX_INIT
config MAINBOARD_FORCE_NATIVE_VGA_INIT
  bool
  depends on MAINBOARD_HAS_NATIVE_VGA_INIT || MAINBOARD_HAS_LIBGFXINIT
choice
  prompt "Graphics initialization"
  config MAINBOARD_DO_NATIVE_VGA_INIT
    bool "Use native graphics init"
    depends on MAINBOARD_HAS_NATIVE_VGA_INIT
  config MAINBOARD_USE_LIBGFXINIT
    bool "Use libgfxinit"
    depends on MAINBOARD_HAS_LIBGFXINIT
  config VGA_ROM_RUN
    bool "Run VGA Option ROMs"
    depends on PCI && !MAINBOARD_FORCE_NATIVE_VGA_INIT
  config NO_GFX_INIT
    bool "None"
    depends on !MAINBOARD_FORCE_NATIVE_VGA_INIT
endchoice
choice
  prompt "Framebuffer mode"
  config VGA_TEXT_FRAMEBUFFER
    bool "Legacy VGA text mode"
    depends on HAVE_VGA_TEXT_FRAMEBUFFER
  config VBE_LINEAR_FRAMEBUFFER
    bool "VESA framebuffer"
    depends on HAVE_VBE_LINEAR_FRAMEBUFFER
endchoice\end{lstlisting}
    \caption{An excerpt from the FM of \textsf{coreboot 4.13}.}
    \label{lst:kconfig-code}
\end{figure}

\begin{figure}[htbp!]
        \centering
        \includegraphics[width=1\linewidth]{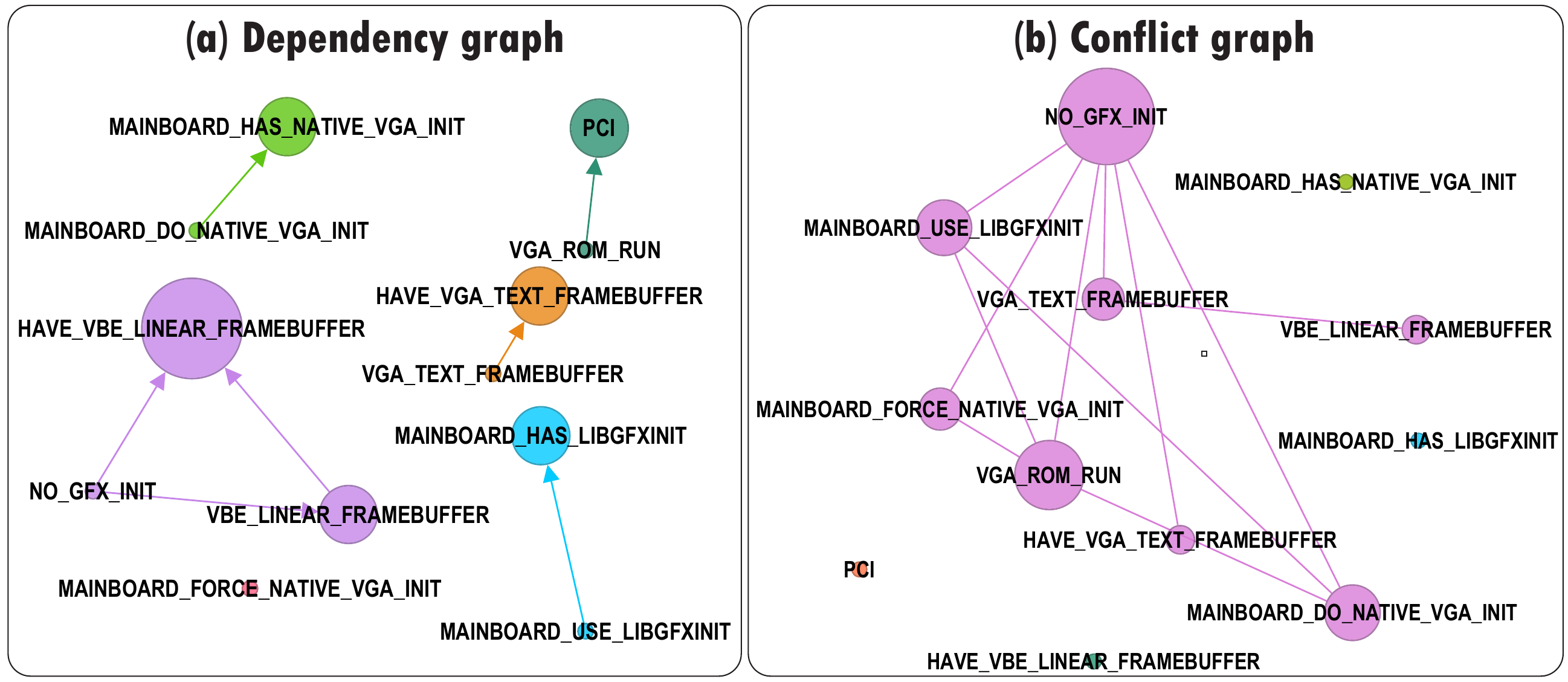}
        \caption{Transitive-closure-based strong graphs for the features in Figure~\ref{lst:kconfig-code}.}
        \label{fig:graph_example}
    \end{figure}

     Figure~\ref{fig:graph_example} shows the strong graphs for Figure~\ref{lst:kconfig-code}. In particular, Figure~\ref{fig:graph_example}.a reveals \textbf{a non-trivial indirect dependency} between \review{\texttt{NO\_GFX\_INIT}} and \texttt{HAVE\_VBE\_} \texttt{LINEAR\_FRAMEBUFFER}, whose explanation reads as follows:

    \begin{enumerate}
        \item \texttt{HAVE\_VGA\_TEXT\_FRAMEBUFFER} excludes \texttt{NO\_GFX\_INIT} (Line 3 in Figure~\ref{lst:kconfig-code}.);
        \item \review{conflicts are symmetric, since $f \Rightarrow \overline{g} \equiv \overline{f} \vee \overline{g} \equiv \overline{g} \vee \overline{f} \equiv g \Rightarrow \overline{f}$. Accordingly, \texttt{NO\_GFX\_INIT} excludes \texttt{HAVE\_VGA\_TEXT\_FRAMEBUFFER}};
        \item as \texttt{VGA\_TEXT\_FRAMEBUFFER} requires \texttt{HAVE\_VGA\_TEXT\_FRAMEBUFFER} (Line 29), disabling \texttt{HAVE\_VGA\_TEXT\_FRAMEBUFFER} produces  that \texttt{VGA\_TEXT\_FRAMEBUFFER} is disabled as well;
        \item as \texttt{VGA\_TEXT\_FRAMEBUFFER} is disabled, and the choice in Lines 22-30 has only two options, the other option must be enabled. So, \texttt{HAVE\_VBE\_LINEAR\_FRAMEBUFFER} must be enabled.
    \end{enumerate}

    The strong graphs also show that:
    \begin{itemize}
        \item the feature with the highest in-degree (i.e., the most depended-upon feature) is \texttt{HAVE\_VBE\_LINEAR\_FRAMEBUFFER};
        \item the feature with the highest out-degree (i.e., the one that depends on the largest number of others) is \texttt{NO\_GFX\_INIT};
        \item \texttt{NO\_GFX\_INIT} and \texttt{VBE\_LINEAR\_FRAMEBUFFER} depend on \texttt{HAVE\_VBE\_LINEAR\_FRAMEBUFFER}, even though this is not obvious from the original Kconfig snippet.
    \end{itemize}

    \newpage{}
    \subsection{Structural analysis of a large FM}\label{sec:large-example}

    As a representative large model, we consider the strong graphs of \textsf{Linux 2.6.9 X86 64-bit}, obtained from translating its Kconfig specification into propositional logic using KConfigReader~\cite{Kuiter25}.
    The resulting formula contains 7,400 Boolean variables\footnote{The Boolean variables represent features, along with auxiliary variables from the Tseitin Boolean translation \cite{Kuiter23}.}
    Based on this formula, we compute the corresponding strong dependency and conflict graphs, identifying core and dead features as described in Section~\ref{sec:how-to}. Simple network-based questions can already yield insights; for example:
    \begin{itemize}
        \item \textbf{What percentage of dead and core features does this model have?}
              In this case, 19.84\% of the features are dead and 5.65\% are core.
        \item \textbf{Which non-core feature is most required by others?}
              The non-core feature with the highest in-degree in the dependency graph is feature \texttt{k!722}, with an in-degree of 1,432 (it is required, directly or indirectly, by features such as \texttt{3C359\_MODULE}, \texttt{PCI}, \texttt{NET\_VENDOR\_3COM}, etc.).
    \end{itemize}

\begin{figure}[htbp!]
        \centering
        \includegraphics[width=0.85\linewidth]{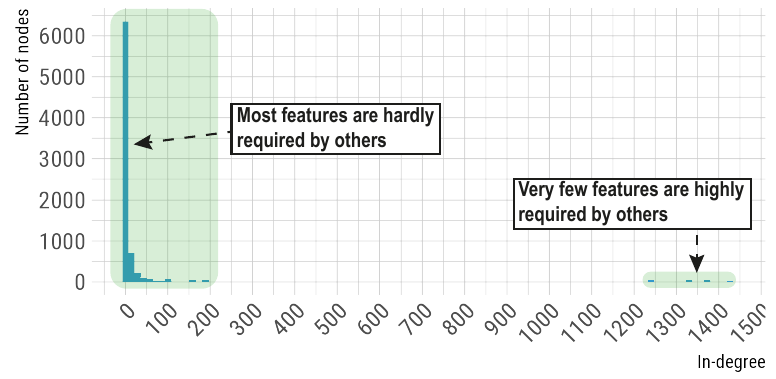}
        \caption{Node in-degree distribution in the dependency graph of Linux 2.6.9 X86 64-bit. Most features are required by few others, while a small number act as highly required hubs.}
        \label{fig:linux-2-6-9-in-degree-dependency}
    \end{figure}

\begin{figure}[htbp!]
        \centering
        \includegraphics[width=0.85\linewidth]{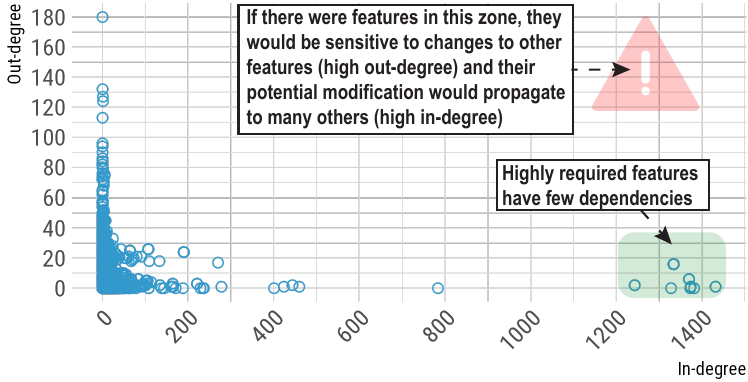}
        \caption{In-degree vs. out-degree in the dependency graph of Linux 2.6.9 X86 64-bit. Highly required features tend to have few dependencies, avoiding features that are both highly required and highly dependent.}
        \label{fig:linux-2-6-9-in-out-degree-dependency}
    \end{figure}

\begin{figure}[htbp!]
        \centering
        \includegraphics[width=0.85\linewidth]{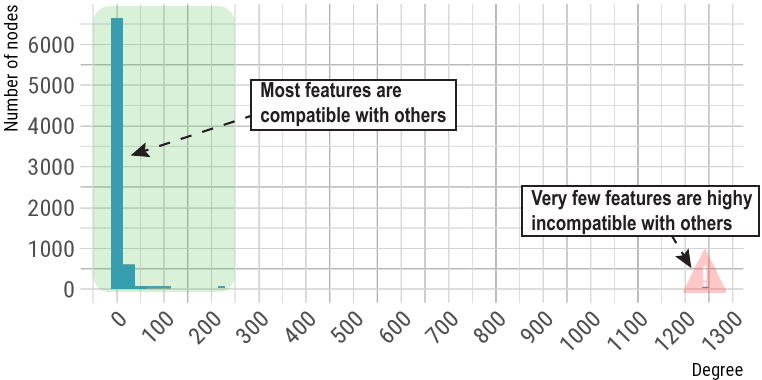}
        \caption{Node degree distribution in the conflict graph of Linux 2.6.9 X86 64-bit. Most features have few conflicts, while a small number of highly constraining features are in conflict with many others.}
        \label{fig:linux-2-6-9-degree-conflict}
    \end{figure}

    To gain a more global view of the model, we look at aggregate network metrics and their distributions.
    Figures~\ref{fig:linux-2-6-9-in-degree-dependency}, \ref{fig:linux-2-6-9-in-out-degree-dependency}, and \ref{fig:linux-2-6-9-degree-conflict} illustrate some of these structural patterns for the \textsf{Linux 2.6.9 X86 64-bit} model.

    Figure~\ref{fig:linux-2-6-9-in-degree-dependency} shows the distribution of node in-degrees in the dependency graph.
    Knowing which features are required by many others is useful for impact analysis, since changes in these features are more likely to affect large parts of the configuration space.
    Most features are hardly required by others (low in-degree), while a small number of features have very high in-degree and act as structural hubs.
    Random changes in most features are therefore unlikely to impact many others, whereas modifications to these hubs may have a disproportionate effect on many dependent features and configurations.

    Figure~\ref{fig:linux-2-6-9-in-out-degree-dependency} relates in-degree and out-degree in the dependency graph.
    Features with many conflicts can be seen as structurally constrained options whose presence excludes many alternative configurations.
    Highly required features tend to have relatively few dependencies themselves, and no feature is both highly required and highly dependent at the same time.
    Features on which many others depend are not, in turn, strongly constrained by large numbers of prerequisites, which suggests a form of structural robustness.

    Finally, Figure~\ref{fig:linux-2-6-9-degree-conflict} shows the distribution of node degrees in the conflict graph.
    Most features are in conflict with relatively few others, and only a small number are incompatible with many alternatives.
    From the point of view of conflicts, selecting a random feature in a configuration is therefore unlikely to exclude a large portion of the feature space.
    The few features with very high conflict degree correspond to highly constraining options whose selection drastically reduces the available configuration choices, and they are natural candidates to inspect when analyzing configurability and potential usability issues in configuration processes.

\vspace{-2mm}
\section{Computing Strong Graphs}
\label{sec:how-to}

In this section, the process applied to compute strong graphs is described. Strong graphs are obtained in two stages.

    \vspace{-2mm}
    \subsection{Stage 1: Translating FMs into Boolean formulas}
    \label{sec:model2bool}

    First, FMs are translated into Boolean formulas~\cite{batory2006automated,Batory05,benavides2010automated,Fernandez19} in \textit{Conjunctive Normal Form} (CNF), which is the required input for most Boolean engines. A CNF is a conjunction of one or more \textit{clauses}, and each clause is a disjunction of \textit{literals}, which can be either a variable or a negated variable.

    \vspace{-2mm}
    \subsection{Stage 2: Computing strong relationships}
    \label{sec:step2}

    Once a FM is encoded as a CNF formula $\varphi$, core/dead features and strong dependency/conflict relations are discovered using the backbone-based Algorithm~\ref {alg:strong-dep}. The \textit{backbone} of a formula is the set of literals with the same truth value in all satisfying assignments of a formula~\cite{Janota2015_Algorithms,Krieter18}.
    Algorithm~\ref {alg:strong-dep} returns (i) a set $B$ with all core and dead features, and (ii) a hashmap $H$ that, for each configurable feature, records the strong relations it has with other features.

    \begin{algorithm}[htbp!]
        \caption{Extraction of strong dependencies and conflicts}
        \begin{flushleft}
            \hspace*{\algorithmicindent} \textbf{Input}: A satisfiable formula $\varphi$ \\
            \hspace*{\algorithmicindent} \textbf{Output}: (i) A set $B$ of all core and dead options; (ii) a hashmap $H$ mapping each configurable option to its strong relations
        \end{flushleft}
        \begin{algorithmic}[1]
            \State $B \gets \mathrm{Backbone}(\varphi)$
            \State $H \gets \emptyset$
            \State $V \gets \operatorname{variables}(\varphi) \setminus \operatorname{variables}(B)$
            \For{$v \in V$}
                \State $B' \gets \mathrm{Backbone}(\varphi \wedge v)$
                \State $H[v] \gets B'$
            \EndFor
            \State remove from $H$ repeated conflict edges
            \State  \Return $(B, H)$
        \end{algorithmic}
        \label{alg:strong-dep}
    \end{algorithm}

    In the beginning, the backbone $B$ of the original formula $\varphi$ is computed.
    Positive literals in $B$ identify core features, while the complements of negative literals identify dead features.
    We then consider all variables $v$ that do not appear in $B$ (i.e., features that are neither core nor dead).
    For each such feature $v$, we compute the backbone $B'$ of the conditioned formula $\varphi \wedge v$.
    Literals in $B'$ describe the consequences of selecting $v$: positive literals represent strong dependencies from $v$ to the corresponding features, and negative literals represent strong conflicts between $v$ and those features.
    These relations are stored in the hashmap $H$, which is then used to create the string graphs.
    Since conflicts are symmetric, duplicate conflict edges are removed at the end (which results is conflicts being represented with undirected exclude edges, while dependencies are represented with directed require arcs).

    We provide a C++ implementation\footnote{\url{https://doi.org/10.5281/zenodo.17790234}} of Algorithm~\ref{alg:strong-dep}, which computes the backbone using the \textit{iterative Algorithm 3} described in~\cite{Janota2015_Algorithms}. Our implementation relies on  \emph{MiniSat 2.2.0}\footnote{\url{http://minisat.se/MiniSat.html}}, which is called through the IPASIR interface\footnote{\url{https://github.com/biotomas/ipasir}} for incremental SAT solving~\cite{Balyo15}.
\vspace{-2mm}
\section{Empirical Network Analysis of FMs}
\label{sec:experimental-validation}

    This section investigates the structural properties of a dataset with 5,709 FMs from open-source and anonymized industrial systems. The FMs were originally published in the 20 academic sources summarized in Table~\ref{tab:dataset-origin}.
    Nevertheless, the starting point for building the strong graphs was not the FMs themselves, but rather the Boolean formulas derived from them by:
    \begin{itemize}
        \item Sundermann et al.~\cite{Sundermann24benchmark} by using the translator TraVarT (\url{https://zenodo.org/records/11654486})~\cite{Feichtinger21}.
        \item Kuiter et al.~\cite{Kuiter25} by using the translators KMax (\url{https://github.com/paulgazz/kmax})~\cite{Oh19tech} and KConfigReader (\url{https://github.com/ckaestne/kconfigreader})~\cite{Kastner17}.
        \item Fernandez et al.~\cite{fernandez23}  by using the tool KconfigSampler (\url{https://github.com/davidfa71/Sampling-the-Linux-kernel})~\cite{fernandez23}.
    \end{itemize}

\begin{table}[htbp!]
\centering
\caption{Number of FMs in the dataset per academic source.}\label{tab:dataset-origin}
\begin{scriptsize}
\begin{tabular}{|c|c|l|r|}
  \hline
\textbf{\#} & \textbf{Ref.} & \textbf{URL of the repository} & \textbf{\#FMs} \\ \hline
  \hline
  1 & \cite{Kuiter2024_SATBasedAnalysisFMs} & \url{https://zenodo.org/records/14884016} & 3,398 \\ \hline
  2 & \cite{Varela-Vaca20} & \url{https://github.com/IDEA-Research-Group/AMADEUS} & 1,463 \\ \hline
  3 & \cite{10.1145/3579027.3608980} & \url{https://github.com/TUBS-ISF/soletta-case-study} & 292 \\ \hline
  4 & \cite{10.1145/3442391.3442410} & \url{https://github.com/TUBS-ISF/SamplingStabiltyVaMoS21\_data} & 248 \\ \hline
  5 & \cite{10.1145/3106237.3106252} & \url{https://github.com/AlexanderKnueppel/is-there-a-mismatch} & 118 \\ \hline
  6 & \cite{Berger13} & \url{https://gsd.uwaterloo.ca/} & 116 \\ \hline
  7 & \cite{10.1145/3579027.3608980} & \url{https://github.com/TUBS-ISF/fiasco-case-study} &  31 \\ \hline
  8 & \cite{10.1145/1639950.1640002} & \url{http://www.splot-research.org/} &  10 \\ \hline
  9 & \cite{10.1145/3393934.3278123} & \url{https://gitlab.com/evolutionexplanation/evolutionexplanation} &  10 \\ \hline
  10 & \cite{oh2020scalable} & \url{https://github.com/jeho-oh/Smarch} &   5 \\ \hline
  11 & \cite{fernandez23} & \url{https://github.com/davidfa71/Sampling-the-Linux-kernel} &   5 \\ \hline
  12 & \cite{khoshmanesh2019leveraging} & \url{https://github.com/zahrakhoshmanesh/FIDUS} &   2 \\ \hline
  13 & \cite{al2019effective} & \url{https://wwwiti.cs.uni-magdeburg.de/iti\_db/research/spl-testing/ }&   2 \\ \hline
  14 & \cite{10.1145/2993236.2993249} & \url{https://wwwiti.cs.uni-magdeburg.de/~jualves/PROFilE/} &   2 \\ \hline
  15 & \cite{10.1145/3336294.3336306} & \url{https://github.com/yamizi/FeatureNet/} &   2 \\ \hline
  16 & \cite{schulze2012variant} & \url{https://github.com/FeatureIDE/FeatureIDE} &   1 \\ \hline
  17 & \cite{kowal2016explaining} & \url{https://github.com/FeatureIDE/FeatureIDE} &   1 \\ \hline
  18 & \cite{hierons2020many} & \url{https://drive.google.com/drive/folders/1xumU6qxBesloq69jOPMbprOaiaOKDq82} &   1 \\ \hline
  19 & \cite{sprey2020smt} & \url{https://github.com/Subaro/SMT-Based-Variability-Analyses-for-FeatureIDE} &   1 \\ \hline
  20 & \cite{lau2006domain} & \url{https://martinfjohansen.com/FMs2011/spltool/} &   1 \\ \hline \hline
  \multicolumn{3}{|r}{\textbf{Total:}} & 5,709 \\ \hline
\end{tabular}
\end{scriptsize}
\end{table}

    In line with our research questions, we aimed to:
    (i) characterize common structural patterns across a large number of FMs (RQ1);
    (ii) analyze how these patterns vary across domains and FM sources (RQ2); and
    (iii) assess to what extent network-based indicators can support practitioners in understanding and maintaining large FMs (RQ3).

    For each formula in the dataset, we computed its strong dependency and strong conflict graphs.
    Both graphs share the same set of nodes, which enables combined interpretations of dependency and conflict metrics.
    We then computed standard network measures and combined them with core/dead information.
    Unless otherwise noted, all analyses were carried out at the level of individual FMs and then aggregated by domain.

    \vspace{-2mm}
    \subsection{Step 1: Graphs Analysis}
    \label{subsec:graphs-analysis}

    For each FM, we followed the steps below:

    \begin{enumerate}
        \item \textbf{Graph construction.}
        We identified the core/dead features and obtained the strong graphs with the procedure described in Section~\ref{sec:how-to}.

        \item \textbf{Structural characterization (RQ1).}
        Using the \textsf{igraph}\footnote{\url{https://igraph.org}} library for network analysis~\cite{Kolaczyk20}, we computed the distributions of node in‑degrees, out‑ degrees, conflict-degrees, and ratios of require vs. exclude links.

        \item \textbf{Domain‑based comparison (RQ2).}
        FMs were grouped by domain.
        For each group, we computed descriptive statistics (i.e., medians and 95\% coverage intervals~\cite{Kaplan12}) together with Spearman's $\rho$ to assess associations between structural metrics and FM size.

        \item \textbf{Derivation of indicators (RQ3).}
        We derived indicators related to maintainability and configurability, focusing on
        highly required and highly conflicting features.
    \end{enumerate}

\begin{figure}[htbp!]
        \centering
        \includegraphics[width=1\linewidth]{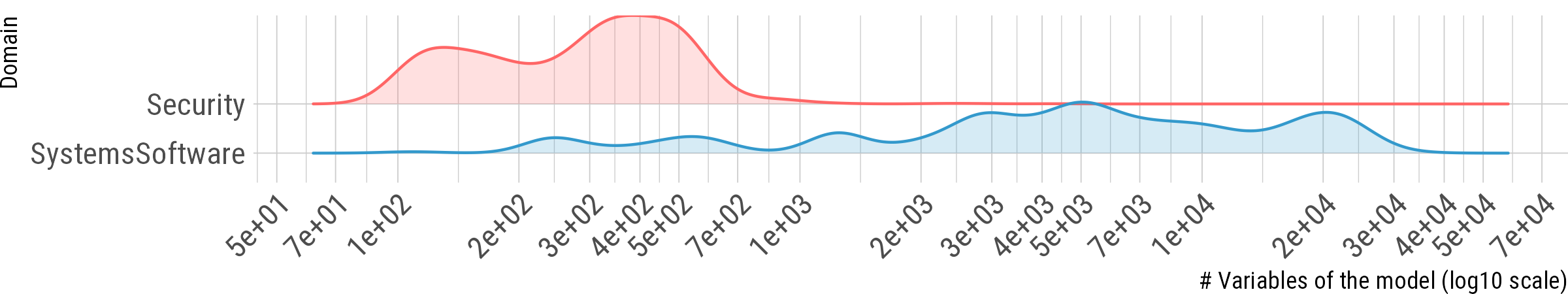}
        \caption{Distribution of the FMs according to their domain and number of variables.}
        \label{fig:density-n-vars}
    \end{figure}

    While the dataset covers several domains (see Table~\ref{tab:core-and-dead-per-domain}), most FMs fall into two categories: \emph{systems software} (4,215 FMs) and \emph{security} (1,464 FMs).
    Figure~\ref{fig:density-n-vars} represents the FM distribution based on the number of variables for both domains. Generally, security models are smaller than systems software models. The size of the FM across the whole dataset ranges from 99 variables (the smallest FM) to 35,907 variables (the largest, which corresponds to a formula generated with KConfigReader from Linux version 4.9).
    The detailed analyses in the following subsections, therefore, focus on these two domains, which together account for the vast majority of the corpus.

    \begin{table}[htbp!]
    \centering
    \caption{Percentage of core and dead variables in the systems per domain.}\label{tab:core-and-dead-per-domain}
    \begin{scriptsize}
    \begin{tabular}{|l|r|c|c|c|c|c|c|}
      \hline
      \multirow{2}{*}{\textbf{Domain}} & \multirow{2}{*}{\textbf{\#FMs}} & \multicolumn{3}{c||}{\textbf{Core variables}} & \multicolumn{3}{c|}{\textbf{Dead variables}} \\ \cline{3-8}
        &  & \textbf{95\% CI } & \textbf{Median} & \textbf{$\rho$} & \textbf{95\% CI} & \textbf{Median} & \textbf{$\rho$} \\
      \hline \hline
      Systems software & 4,215 & 0.00\%-41.09\% & 5.30\% & 0.59 & 0.00\%-42.35\% & 18.27\% & 0.41 \\ \hline
      Security & 1,464 & 0.17\%-10.04\% & 2.31\% & -0.84 & 0.00\%-98.45\% & 0.00\% & 0.58 \\ \hline
      Finance &  13 & 2.42\%-9.51\% & 2.85\% & -0.49 & 0.00\%-0.48\% & 0.00\% & 0.60 \\ \hline
      Automotive &   5 & 4.48\%-9.91\% & 9.55\% & 0.30 & 0.05\%-6.99\% & 0.05\% & -0.70 \\ \hline
      Deep learning &   2 & 0.02\%-0.21\% & 0.11\% & $-$ & 0.04\%-1.42\% & 0.73\% & $-$ \\ \hline
      eCommerce &   2 & 0.77\%-28.18\% & 14.47\% & $-$ & 0.00\%-0.00\% & 0.00\% & $-$ \\ \hline
      Hardware &   2 & 3.04\%-27.84\% & 15.44\% & $-$ & 0.00\%-0.00\% & 0.00\% & $-$ \\ \hline
      Navigation &   2 & 17.18\%-41.12\% & 29.15\% & $-$ & 0.00\%-\%0.00 & 0.00\% & $-$ \\ \hline
      Business &   1 & $-$ & 0.05 & $-$ & $-$ & 0.00 & $-$ \\ \hline
      Database &   1 & $-$ & 11.97\% & $-$ & $-$ & 5.13 & $-$ \\ \hline
      Games &   1 & $-$ & 5.56\% & $-$ & $-$ & 0.00\% & $-$  \\ \hline
      Text &   1 & $-$ & 5.11\% & $-$ & $-$ & 0.00\% & $-$ \\ \hline
    \end{tabular}
    \end{scriptsize}
    \end{table}

    \vspace{-1cm}

    \vspace{-2mm}
    \subsection{Step 2: Core and dead features across domains}
    \label{subsec:core-dead}

    Table~\ref{tab:core-and-dead-per-domain} summarizes the proportion of core and dead variables per FM in each domain.
    For each combination (domain, variable type) we report:
    (i) the 95\% coverage interval (CI) (i.e., excluding the 2.5\% lowest and 2.5\% highest),
    (ii) the median percentage across FMs, and
    (iii) Spearman's $\rho$ between the percentage and the number of variables in the FM.

    For example, the first row indicates that in the systems-software domain, 95\% of the FMs have between 0\% and 41.09\% of their core variables.
    The median is 5.30\%, so half of these FMs have a core ratio below that value and half above.
    The correlation $\rho=0.59$ suggests that larger systems-software FMs tend to exhibit a higher proportion of core features.

    For security FMs, core features are relatively scarce (median 2.31\%), and the negative correlation ($\rho=-0.84$) indicates that larger FMs in this domain tend to have even fewer core features in percentage terms.
    Conversely, dead features are extremely concentrated in some security FMs (the CI upper bound is close to 100\%), yet the median is 0\%, reflecting a very skewed distribution: many FMs have virtually no dead features, while a few have highly constrained outliers.
    This contrasts with the systems-software domain, where dead features are more common (median 18.27\%).

    To statistically compare core and dead features, Table~\ref{tab:wilcoxon-core-and-dead-per-domain} reports Wilcoxon signed-rank tests per domain.
    In the systems software domain, FMs tend to contain significantly more dead variables than core variables (moderate effect size\footnote{Effect size was computed as $r=\frac{Z}{\sqrt{N}}$ \cite{Field12}, where $Z$ is the standardized test statistic and $N$ is the sample size. The common interpretation is $|r|<0.3$ (small effect size), $0.3\leq |r| < 0.5$ (moderate effect), and $|r| \geq 0.5$ (large effect).}), whereas in security, the difference is not statistically significant.

    \begin{table}[htbp!]
    \centering
    \caption{Wilcoxon signed-rank test to check if the FMs have more dead variables than core variables.}\label{tab:wilcoxon-core-and-dead-per-domain}
    \begin{scriptsize}
    \begin{tabular}{|l||c||r|c||r|c|}
      \hline
      \textbf{Domain} & \textbf{Alt. Hyp. ($H_a$)} & \textbf{$p$-value} & \textbf{Stat. sig.?} & \textbf{$r$} & \textbf{(effect size)} \\ \hline
      Systems software & \#Dead $>$ \#Core & < 2.2e-16 & Yes & 0.38 & Moderate \\ \hline
      Security & \#Dead $>$ \#Core & 0.5 & No & 0.04 & Negligible \\ \hline
    \end{tabular}
    \end{scriptsize}
    \end{table}

    Overall, these results already suggest that structural behavior is far from uniform across domains.

    \vspace{-2mm}
    \subsection{Step 3: Dependencies versus conflicts}
    \label{subsec:req-excl}

    Table~\ref{tab:requires-and-excludes-per-domain} analyzes the density of strong require and exclude relations.
    For each domain, we normalized the number of require arcs and conflict edges by the number of variables in the FM formula. We again report 95\% CIs, medians, and Spearman's $\rho$.

  \begin{table}[htbp!]
    \caption{Proportion of require and exclude links relative to the number of variables in the systems per domain.}
    \label{tab:requires-and-excludes-per-domain}
    \centering
    \begin{scriptsize}
    \begin{tabular}{|l|r||c|c|c||c|c|c|}
      \hline
      \multirow{2}{*}{\textbf{Domain}} & \multirow{2}{*}{\textbf{\#FMs}} & \multicolumn{3}{c||}{\textbf{Require arcs}} & \multicolumn{3}{c|}{\textbf{Exclude edges}} \\ \cline{3-8}
        &  & \textbf{95\% CI } & \textbf{Median} & \textbf{$\rho$}  & \textbf{95\% CI} & \textbf{Median} & \textbf{$\rho$} \\
      \hline \hline
        Systems software & 4,215 & 1.12$\times$-75.95$\times$ & 7.11$\times$ & 0.47 & 0.09$\times$-49.18$\times$ & 2.78$\times$ & 0.37 \\ \hline
        Security & 1,464 & 0.00$\times$-7.16$\times$ & 0.82$\times$ & -0.01 & 0.04$\times$-259.24$\times$ & 62.65$\times$ & 0.02 \\ \hline
        Finance &  13 & 0.58$\times$-49.46$\times$ & 34.67$\times$ & 0.92 & 0.12$\times$-272.08$\times$ & 259.37$\times$ & 0.43 \\ \hline
       Automotive &   5 & 1.28$\times$-30.92$\times$ & 1.32$\times$ & -0.40 & 8.14$\times$-18.13$\times$ & 17.69$\times$ & 0.60 \\ \hline
       Deep learning &   2 & 14.66$\times$-24.21$\times$ & 19.44$\times$ & $-$ & 5.46$\times$-14.11$\times$ & 9.78$\times$ & $-$ \\ \hline
       eCommerce &   2 & 4.95$\times$-10.53$\times$ & 7.74$\times$ & $-$ & 0.00$\times$-0.00$\times$ & 0.00$\times$ & $-$ \\ \hline
       Hardware &   2 & 0.22$\times$-1.16$\times$ & 0.69$\times$ & $-$ & 1.97$\times$-16.59$\times$ & 9.28$\times$ & $-$ \\ \hline
       Navigation &   2 & 0.37$\times$-4.91$\times$ & 2.64$\times$ & $-$ & 0.23$\times$-5.17$\times$ & 2.70$\times$ & $-$ \\ \hline
       Business &   1 & $-$ & 94.62$\times$ & $-$ & $-$ & 0.00$\times$ & $-$ \\ \hline
       Database &   1 & $-$ & 73.13$\times$ & $-$ & $-$ & 0.00$\times$ &  $-$\\ \hline
       Games &   1 & $-$ & 0.50$\times$ & $-$ & $-$ & 4.22$\times$ & $-$ \\ \hline
       Text &   1 & $-$ & 3.26$\times$ & $-$ & $-$ & 0.00$\times$ &  $-$\\
       \hline
    \end{tabular}
    \end{scriptsize}
    \end{table}

        In systems software, require relations are much more common than conflicts. On average, there are 7.11 times more require arcs than nodes, compared to 2.78 times more exclude edges. Both of these metrics increase with the size of the FM.

    In contrast, security FMs shows very dense conflict graphs (median 62.65$\times$) and comparatively sparse dependency structures (median 0.82$\times$).
    Wilcoxon tests in Table~\ref{tab:wilcoxon-excludes-and-requires-per-domain} confirm that:
    (i) in systems software, require arcs are significantly predominant (large effect size), whereas
    (ii) in security, exclude edges significantly dominate the require arcs.

    \begin{table}[htbp!]
    \centering
    \caption{Wilcoxon signed-rank test to check if the FMs have more require arcs or exclude edges.}\label{tab:wilcoxon-excludes-and-requires-per-domain}
    \begin{scriptsize}
    \begin{tabular}{|l||c||r|c||r|c|}
      \hline
      \textbf{Domain} & \textbf{Alt. Hyp. ($H_a$)} & \textbf{$p$-value} & \textbf{Stat. sig.?} & \textbf{$r$} & \textbf{(effect size)} \\ \hline
      Systems software & \#Excludes $<$ \#Requires & < 2.2e-16 & Yes & -0.67 & Large \\ \hline
      Security & \#Excludes $>$ \#Requires & < 2.2e-16 & Yes & 0.87 & Large \\ \hline
    \end{tabular}
    \end{scriptsize}
    \end{table}


    \vspace{-2mm}
    \subsection{Step 4: Degree distributions and hubs}
    \label{subsec:degree-hubs}

\begin{figure}[htbp!]
        \centering
        \includegraphics[width=0.95\linewidth]{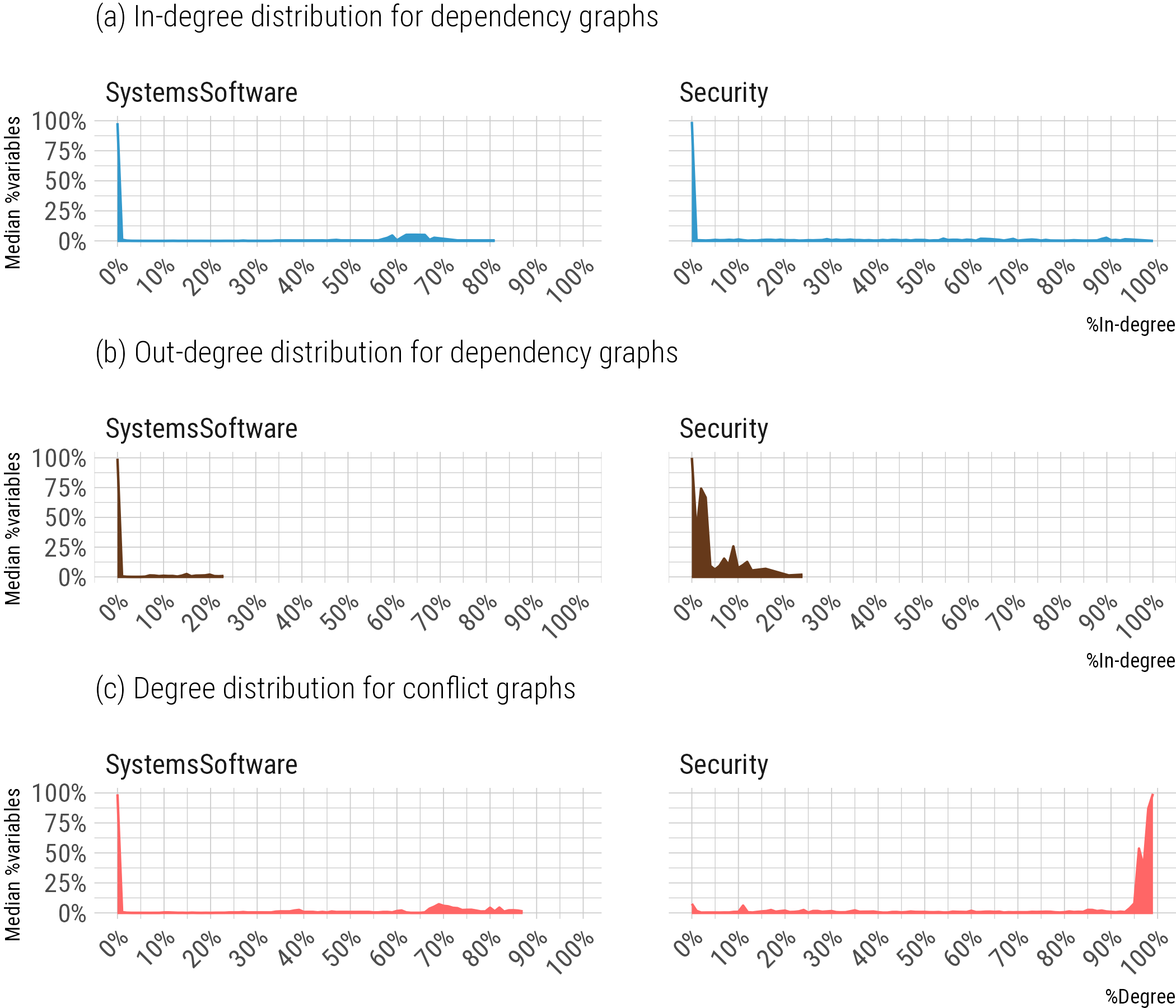}
        \caption{Node degree distributions across FMs (dependency in-/out-degree and conflict degree).}
        \label{fig:degree-distributions}
    \end{figure}

    To further understand structural organization, Figure~\ref{fig:degree-distributions} shows degree distributions across FMs.
    The $x$-axis represents node degrees as percentages of other features in the same FM, and the $y$-axis shows, for each degree value, the median percentage of features that exhibit that degree.

    We note that a characteristic pattern emerges.
    Most nodes exhibit very low in-degree and conflict degree, while a small fraction of nodes accumulates very high values.
    For instance, in the systems-software dependency graphs, there is a pronounced peak where, on median, around 8\% of the nodes depend on more than half of the other features in the FM.
    Similarly, conflict-degree distributions reveal that only a small subset of features have conflicts with a large portion of the feature space.

    This confirms the existence of \emph{configuration hubs}:
    features with exceptionally high in-degree (strongly required by many others) and, in some domains, features with exceptionally high conflict degree (highly constraining options).
    These hubs are potential weak points:
    changes or threats affecting them may propagate to a large number of dependent or incompatible options.

\begin{figure}[htbp!]
        \centering
        \includegraphics[width=0.8\linewidth]{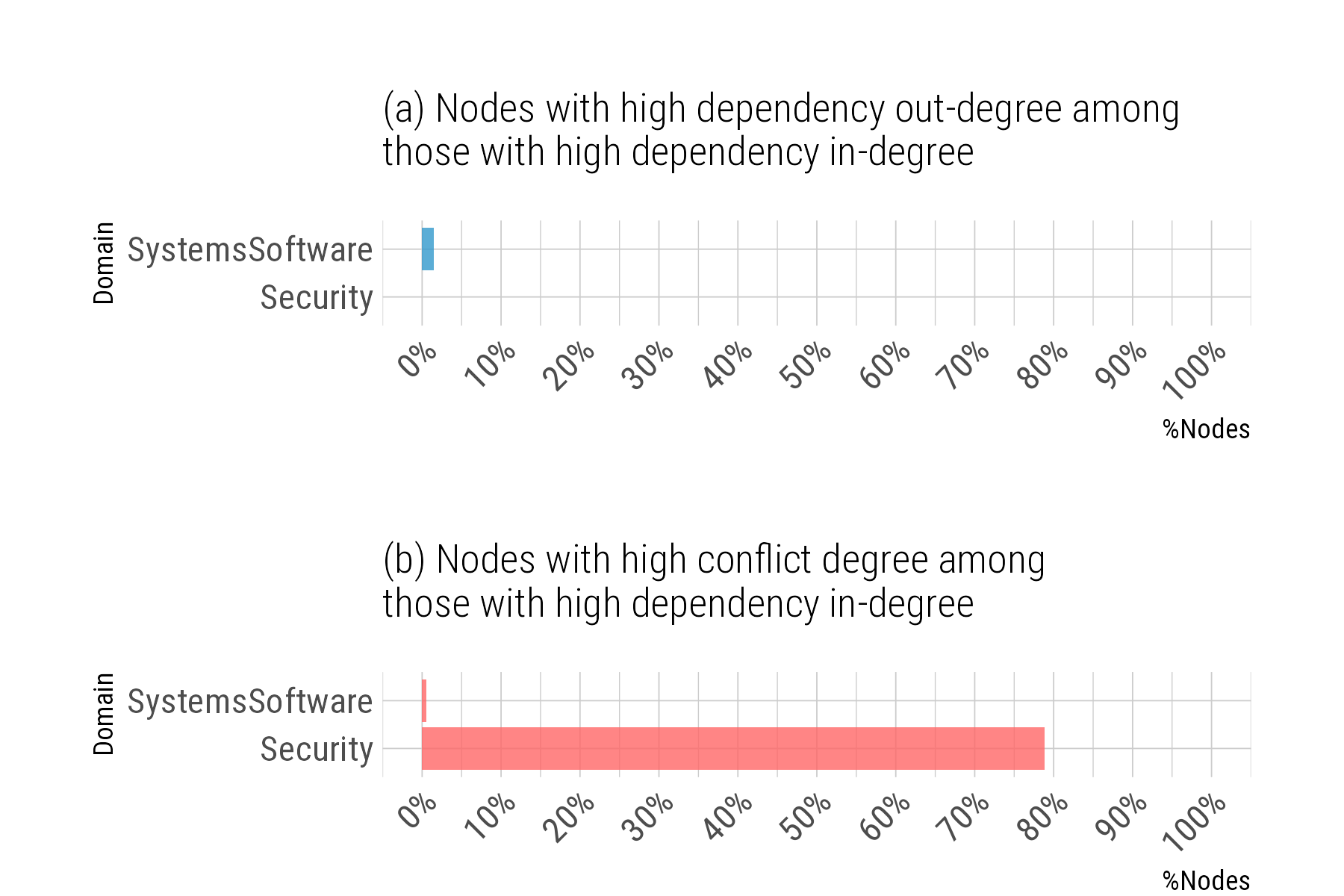}
        \caption{Conditional distributions of nodes with high degrees (degree $\geq$ 10\% of \#nodes).}
        \label{fig:degree-overlapping}
    \end{figure}

    Figure~\ref{fig:degree-overlapping} complements this analysis by examining how high degrees overlap.
    We consider a degree to be \emph{high} when it is at least 10\% of the features in a FM (e.g., if a node is required by more than 10\% of the remaining nodes, its in-degree is high); this threshold could be changed to adjust the sensitivity of the analysis.

    Figure~\ref{fig:degree-overlapping}.a relates in-degree and out-degree in dependency graphs.
    High percentages in this panel indicate the presence of features in the “warning zone” highlighted in Figure~\ref{fig:linux-2-6-9-in-out-degree-dependency}, (i.e., features that are both highly required and highly dependent).
    Such features are problematic from a maintainability perspective, as they are central in both directions: they affect many others and are themselves constrained by many prerequisites.

    Figure~\ref{fig:degree-overlapping}.b relates dependency in-degree and conflict degree.
    For security FMs, almost 80\% of nodes with high dependency in-degree also have high conflict degree.
    This indicates that highly required features in these FMs tend to be largely incompatible with many alternatives, suggesting limited reuse and fragmentation into relatively isolated subsystems.

    These observations instantiate the second pattern mentioned in the motivational discussion in Section~\ref{sec:large-example}:
    across FMs, there is a clear tendency for a minority of features to act as hubs (in dependencies and, in some domains, in conflicts), while most features remain structurally peripheral.

    \vspace{-2mm}

    \vspace{-2mm}
    \subsection{Discussion}
    \label{subsec:discussion}

    The empirical evidence shows that strong graph analysis scales to FMs with thousands of features and reveals:
    (i) stable structural patterns (RQ1),
    (ii) clear differences across domains and FM sources (RQ2), and
    (iii) practical indicators for spotting critical features and substructures (RQ3).
    These findings are not apparent from the original variability artifacts or their raw Boolean encodings alone, highlighting the \review{additional descriptive insight} of a network-based perspective for both researchers and practitioners.

    Taken as a whole, the empirical results outline a coherent picture of how FMs are organized structurally, which can be interpreted in terms of the research questions proposed in Section~\ref{sec:introduction}.

    \paragraph*{RQ1: Structural patterns in strong graphs.}
    Across the 5,709 FMs in our dataset, we see several recurring structural patterns.
    First, in systems-software FMs, strong dependencies are much more frequent than strong conflicts (Table~\ref{tab:requires-and-excludes-per-domain}), and their density \review{tends to increase} with FM size.
    Second, degree distributions in both dependency and conflict graphs are highly uneven (Figure~\ref{fig:degree-distributions}).

    \review{These results suggest that FMs are not structurally flat or random: they exhibit recognizable, repeated shapes in their strong graphs.}

    \paragraph*{RQ2: Variation across domains and sources.}
    The structural behavior of FMs changes across domains and FM sources.
    Systems-software FMs (e.g., Linux, BusyBox, and related systems) have moderate but consistent proportions of core and dead features (Table~\ref{tab:core-and-dead-per-domain}), and larger FMs tend to have more core and dead options in a controlled way.
    Their strong graphs are dense in dependencies and comparatively light in conflicts (Table~\ref{tab:requires-and-excludes-per-domain}).
    In contrast, many security FMs, which are often \review{synthetically} generated from vulnerability repositories, show very high conflict densities and very skewed distributions of dead features: most FMs have almost no dead features, but a few contain very high percentages.
    Figure~\ref{fig:degree-overlapping} also shows that, in security FMs, highly required features often coincide with highly conflicting ones, which \review{may indicate} structurally fragile areas where central features are strongly constrained.
    Smaller domains (finance, automotive, navigation, etc.) show their own characteristic profiles, but the low number of FMs means that we must be careful with strong claims.
    Consequently, \review{our results do not support} single structural pattern that fits all domains, they suggest domain- and source-dependent structural tendencies.

    \paragraph*{RQ3: Usefulness of network-based indicators.}
    The structural indicators derived from strong dependency and conflict graphs \review{provide a compact way to describe feature-level and substructure-level positions in the analyzed FMs.
    While these indicators appear potentially relevant for maintenance and evolution, such implications remain exploratory in our study because we did not directly measure maintenance outcomes.}
    \review{(i)  \textit{Features with very high dependency in-degree} may indicate structurally central elements that could become maintenance-sensitive areas, although this requires validation with direct maintainability evidence.
    }
    (ii) \textit{Features with high conflict degree} are \emph{\review{configurationally restrictive}}: selecting them removes many alternatives from the configuration space and may create problems during configuration, especially in domains with dense conflict graphs such as security.
    Finally, (iii) \textit{the overlap between high dependency and high conflict} degrees (Figure~\ref{fig:degree-overlapping}) indicates that highly required features in some FMs are often incompatible with many other features, \review{which may be associated with} limited reuse and fragmentation into relatively isolated subsystems.

    \review{
    The observed differences between systems-software and security FMs should also be interpreted in light of different domains' characteristics and models' origin.
    In our dataset, many systems-software FMs were derived from manually curated models of system capabilities, whereas many security FMs were synthetically generated with computer-assisted procedures from vulnerability repositories.
    Therefore, some of the structural differences may reflect not only domain-specific variability but also differences in modeling process and artifact construction.
    Future work should compare alternative FMs for the same systems (e.g., manually produced vs. computer-assisted generated) to assess how much the observed graph properties depend on the modeling methodology.}

    \vspace{-2mm}
    \subsection{Threats to validity}
    \label{sec:threats}

    Several factors may affect the validity of the reported results.

    \textbf{Construct validity.}
    Strong relationships represent only a subset of the relations that exist in an FM. Our analysis focuses exclusively on strong relationships; those semantic dependencies and conflicts that hold across every valid configuration of the FM. While this provides a conservative and well-defined foundation for analysis, it means we do not capture weaker forms of correlation or probabilistic associations between features that may occur in many (but not all) valid configurations. This choice trades breadth for rigor: by focusing on relations guaranteed to hold universally, we obtain stable, verifiable structural patterns, but we necessarily omit more nuanced relationships that might be relevant in certain contexts.

    Additionally, the quality of our analysis depends critically on the translation from FM to Boolean formulas. For FMs expressed in simple variability languages such as the \textit{Universal Variability Language} (UVL)~\cite{benavides2024uvl}, where semantics are well-defined and standardized, this translation is straightforward and unproblematic. However, for Kconfig FMs, which constitute a substantial portion of our dataset (Linux, Busybox, and related systems), the situation is more complex. Despite numerous efforts to translate Kconfig into Boolean logic \cite{Berger13,Fernandez19,Kastner17,Sincero10under,Yaman24}, none have achieved completeness as they do not fully adhere to the Kconfig syntax, and none have been thoroughly validated \cite{Fernandez19}. Consequently, the Boolean formulas derived from Kconfig specifications may not fully capture the original semantics, and the strong graphs computed from these formulas may underestimate or misrepresent certain dependencies and conflicts.

    \textbf{Internal validity.}
    The implementation addresses potential errors from two main sources: backbone computation and core/dead option identification ($S_1$), and graph construction ($S_2$). To mitigate $S_1$, we cross-validated backbone results using an independent tool (MiniBones). For $S_2$, we implemented a validation algorithm\footnote{The validation algorithm is available at \url{https://doi.org/10.5281/zenodo.17790233}} that systematically verifies correctness by testing all core and dead options, and then randomly sampling graph nodes to confirm that dependency and conflict relationships are accurate.

    The validation process was executed comprehensively across all 5,709 models, testing approximately 5.7 million nodes over two weeks using parallel processing. The extensive computational effort, examining 1,000 nodes per model on high-performance hardware, with zero errors detected across all tested arcs and edges provides empirical evidence that the dataset and graph construction are correct. This rigorous validation approach significantly strengthens confidence in the accuracy of the subsequent structural analysis findings.

    \textbf{External validity.}
    The dataset is limited to publicly available FMs from specific ecosystems, and some domains are represented by only a few FMs.
    As a result, domain-specific observations for small groups should be regarded as indicative rather than conclusive.

    \textbf{Conclusion validity.}
    Our statistical conclusions depend on appropriate analysis methods and sound interpretation of results. Several factors could affect the validity of our conclusions. First, structural metrics depend on FM size; we address this by computing correlations between metrics and FM size to account for size effects. Second, FMs from the same project may not be fully independent, potentially leading to inflated statistical significance. To mitigate this, we used non-parametric statistical tests (Spearman's $\rho$ and Wilcoxon signed-rank tests) that make minimal assumptions about data distribution. Third, we report effect sizes alongside p-values to assess practical significance. Finally, different FM-to-Boolean translators may produce different results, introducing variation in graph metrics. Further replications on broader datasets would strengthen confidence in our conclusions.

\vspace{-2mm}
\section{Conclusions}
\label{sec:conclusions}

This work describes how to create strong graphs from variability models and demonstrates that these graphs provide a foundation for understanding their structural organization.
By analyzing 5,709 FMs across different domains this study establishes network analysis as a viable and insightful complement to logical reasoning in variability modeling, enabling practitioners and researchers to systematically understand how variability is organized across domains.
We reveal both universal structural patterns and domain-specific aspects.
The following paragraphs outline the main advantages of strong graphs that this paper emphasizes:

    \textbf{Strong graphs enable fine-grained feature analysis}. They reveal subtle relationships arising from the interplay of constraints, surfacing hidden dependencies and conflicts that manual inspection misses. Strong graphs quantify each feature's role: how many others depend on it, how many it depends on, and how many it conflicts with. This supports targeted impact analysis and helps practitioners prioritize testing and code review efforts.

    \textbf{Strong graphs enable holistic FM assessment}. At the system level, they reveal whether a FM exhibits structural robustness, where highly required features have few dependencies, or fragility, where central features are both highly required and highly dependent. Strong graphs also expose whether a system is compact or fragmented into isolated subsystems with limited reuse.

    \textbf{Strong graphs enable empirical characterization across domains}. Systems software FMs exhibit a dependency-dominated organization with robust hub-and-spoke patterns. Security FMs, by contrast, sometimes have extremely dense conflict graphs and are fragmented into subsystems. These domain-specific patterns reflect both problem constraints and engineering practices, helping practitioners benchmark their models and adopt appropriate refactoring strategies.
    \review{Future work includes analyzing similarities and differences of software systems being described as a FM, either manually by examining their features or automatically (e.g., considering their requirements).}

    \textbf{Practical implications}. Configuration tool developers can leverage degree-based metrics for intelligent user guidance. Maintainers can identify refactoring priorities by locating features in the structural ``warning zones'' (Figures~\ref{fig:linux-2-6-9-in-out-degree-dependency} and \ref{fig:linux-2-6-9-degree-conflict}). Testing practitioners can focus their efforts on high-impact features through impact-aware prioritization.

\vspace{-2mm}

\section*{Materials}

In line with open science principles, the software artifacts needed to replicate our experimental validation in Section~\ref{sec:experimental-validation} are available in the following public repository: \url{https://doi.org/10.5281/zenodo.17790233}


\vspace{-2mm}

\begin{credits}
\subsubsection{\ackname}

This work is funded by FEDER/Spanish Ministry of Science, Innovation and Universities (MCIN)/Agencia Estatal de Investigacion (AEI) under grant codes COSY (PID2022-142043NB-I00) and Data-PL (PID2022-138486OB-I00), and by Junta~de~Andalucía under grant code SENSOLIVE (PLSQ\_00162).

\subsubsection{\discintname}
The authors have no competing interests to declare that are relevant to the content of this article.
\end{credits}

\bibliographystyle{splncs04}
\bibliography{references}

\end{document}